\documentclass[11pt]{article}
\usepackage{graphicx}
\usepackage{rotating}

\newcommand{\BABARPubYear}    {01}

\newcommand{\BABARConfNumber} {10}

\newcommand{\LANLNumber} {0107058}

\input pubboard/babarsym

\newcommand{\BrRhoppim}{ 28.9 \pm  5.4 \pm  4.3}
\newcommand{\AsRhoppim}{-0.04 \pm 0.18 \pm 0.02}
\newcommand{\BrRhozpiz}{3.6 \pm  3.5 \pm  1.7}
\newcommand{\BrRhopr}{ 5.1 \pm  2.9 \pm  2.2}
\newcommand{\BrSigmapm}{ 2.5 \pm  2.1 \pm  0.8}

\newcommand{\BrRhoppimVal}{(\BrRhoppim)\times 10^{-6}}
\newcommand{\AsRhoppimVal}{\AsRhoppim}

\newcommand{\UlRhozpiz}{10.6}
\newcommand{\UlRhopr}{11.3}
\newcommand{\UlRhoprz}{2.7}
\newcommand{\UlSigmapm}{6.1}
\newcommand{\UlSigmaz}{5.2}

\newcommand{\UlPipipi}{7.3}

\newcommand{\UlPipipiVal}{\UlPipipi\times 10^{-6}}

\newcommand{\rhoppim}{\mbox{$\Bz \to \rho^{\mp} \pi^{\pm}$}}
\newcommand{\rhozpiz}{\mbox{$\Bz \to \rho^0 \pi^0$}}
\newcommand{\rhopr}{\mbox{$\Bz \to \rho '^{\mp} \pi^{\pm}$}}
\newcommand{\rhoprz}{\mbox{$\Bz \to \rho '^{0} \pi^{0}$}}
\newcommand{\sigmapm}{\mbox{$\Bz \to {\rm~charged~scalar~and~} \pi^{\mp}$}}
\newcommand{\sigmaz}{\mbox{$\Bz \to {\rm~neutral~scalar~and~} \pi^0$}}
\newcommand{\highmass}{\mbox{$\Bz \to \pi^+ \pi^- \pi^0$ at high mass}}

\newcommand{\pipipiz}{\mbox{$\pi^+ \pi^- \pi^0$}}

\newcommand{\pppz}{$\Bz \to \pip \pim \piz$}

\newcommand{\DE}{\ensuremath{\Delta E}}
\newcommand{\de}{\ensuremath{\Delta E}}
\newcommand{\calB}{\mbox{${\cal B}$}}

\newcommand{\xf}{\mbox{${\cal F}$}}
\newcommand{\sigGSBratio}{\calR}
\newcommand{\half}{\mbox{${1\over2}$}}
\newcommand{\pvec}{{\bf p}}

\newcommand{\ie}{{\it ie.}}

\def\dbline{\noalign{\vskip 0.15truecm\hrule}\noalign{\vskip 2pt}\noalign{\hrule\vskip 0.15truecm}}
\def\sgline{\noalign{\vskip 0.20truecm\hrule\vskip 0.20truecm}}

\def\Y#1S{{\Upsilon\rm(#1S)}}

\def\ra{\rightarrow}
\def\etal{{\it et~al.,}}

\long\def\inst#1{\par\nobreak\kern 4pt\nobreak
    {\it #1}\par\vskip 10pt plus 3pt minus 3pt}

\begin{document}
{\pagestyle{empty}

\begin{flushright}
\babar-CONF-\BABARPubYear/\BABARConfNumber \\
hep-ex/\LANLNumber \\
July, 2001 \\
\end{flushright}

\par\vskip 4cm

\begin{center}
\Large \bf Measurements of $B^0$ decays to $\pi^+\pi^-\pi^0$
\end{center}
\bigskip

\begin{center}
\large The \babar\ Collaboration\\
\mbox{ }\\
July 21, 2001
\end{center}
\bigskip \bigskip

\begin{center}
\large \bf Abstract
\end{center}
We present preliminary results of 
searches for exclusive $B^0$ decays to $\pi^+\pi^-\pi^0$ among 22.7 million
$b\overline{b}$ pairs collected by the \babar\ experiment from
electron-positron collisions near the $\FourS$ resonance. 
We measure $\calB(\rhoppim) = \BrRhoppimVal$, and find no evidence for 
the presence of any other decay mode in the $\pi^+\pi^-\pi^0$ Dalitz plot.
Upper limits are determined for the branching fractions of
$\rhozpiz$, non-resonant 
$B^0$ decays to $\pi^+\pi^-\pi^0$ and of several discrete regions of 
$\pi^+\pi^-\pi^0$ phase-space.
We also measure the direct $CP$-violating asymmetry between the rates of 
untagged $\rho^+\pi^-$ and $\rho^-\pi^+$, finding no significant evidence 
for an effect.

\vfill
\begin{center}
Submitted to the\\ 20$^{th}$ International Symposium 
on Lepton and Photon Interactions at High Energies, \\
7/23---7/28/2001, Rome, Italy
\end{center}

\vspace{1.0cm}
\begin{center}
{\em Stanford Linear Accelerator Center, Stanford University, 
Stanford, CA 94309} \\ \vspace{0.1cm}\hrule\vspace{0.1cm}
Work supported in part by Department of Energy contract DE-AC03-76SF00515.
\end{center}
}

\newpage

\begin{center}
\small

The \babar\ Collaboration,
\bigskip

B.~Aubert,
D.~Boutigny,
J.-M.~Gaillard,
A.~Hicheur,
Y.~Karyotakis,
J.~P.~Lees,
P.~Robbe,
V.~Tisserand
\inst{Laboratoire de Physique des Particules, F-74941 Annecy-le-Vieux, France }
A.~Palano
\inst{Universit\`a di Bari, Dipartimento di Fisica and INFN, I-70126 Bari, Italy }
G.~P.~Chen,
J.~C.~Chen,
N.~D.~Qi,
G.~Rong,
P.~Wang,
Y.~S.~Zhu
\inst{Institute of High Energy Physics, Beijing 100039, China }
% ...

\end{center}\newpage

\setcounter{footnote}{0}

\section{Introduction}
\label{sec:Introduction}

We describe the measurement of branching fractions for 
$B^0$ decays\footnote{Charge conjugate initial and final states are assumed 
everywhere unless otherwise stated.} to the final state \pipipiz.
We include here those decays which proceed through a resonant, quasi-two body
intermediate state, as well as non-resonant three-body decays with which they 
interfere. 

The intrinsic interest in these modes stems from their potential use 
for measuring the angle $\alpha$ of the unitarity triangle \cite{book}. 
Such measurements would exploit interference between the \rhoppim\ modes 
and the colour-suppressed \rhozpiz\ mode. However, the former modes were 
discovered only in 1999 \cite{cleo2000}, and the latter mode remains 
undiscovered. Furthermore, the proposed measurement of $\alpha$ assumes 
that only the lowest-mass $\rho$ mesons contribute to the \pipipiz\ 
Dalitz plot, an assumption that needs to be tested. It has also been
suggested \cite{Bstar, sigma} that resonances with quantum numbers 
other than those of the $\rho$ might make significant contributions 
to the rate in the Dalitz plot, and it is important also to search for these.
In the analyses presented here, we improve on earlier measurements of 
\rhoppim\ and search for \rhozpiz\ modes containing other resonances, 
and non-resonant \pppz\ decays.

\section{The \babar\ Detector and Dataset}
\label{sec:babar}
The data used in these analyses were collected with the \babar\ detector
at the \pep2\ storage ring. 
The \babar\ detector, described in detail elsewhere~\cite{ref:babar}, 
consists of five active sub-detectors. Surrounding the beam-pipe is a silicon 
vertex tracker (SVT) to track particles of momentum less than $\sim$120\mevc\ 
and to provide precision measurements of the positions of charged
particles of all momenta as they leave the interaction point. A beam-support
tube surrounds the SVT. Outside this is a 40-layer drift
chamber (DCH), filled with an 80:20 helium-isobutane gas mixture to
minimize multiple scattering, providing measurements of track momenta
in a 1.5 T magnetic field.  It also provides energy-loss measurements  
that contribute to charged particle identification. Surrounding the outer
circumference of the drift chamber is a novel detector of internally
reflected Cherenkov radiation (DIRC) that provides charged hadron
identification in the barrel region. This consists of quartz bars of
refractive index $\sim$1.42 in which Cherenkov light is produced by
relativistic charged particles. This is internally reflected and
collected by an array of photomultiplier tubes, which enable Cherenkov
rings to be reconstructed and associated with the charged tracks in
the DCH, providing a measurement of particle velocities. Outside
the DIRC is a CsI(Tl) electromagnetic calorimeter (EMC) which is used to
detect photons and neutral hadrons, and to provide electron identification. 
The EMC is surrounded by a superconducting coil which provides the magnetic 
field for tracking. Outside the coil, the flux return is instrumented with
resistive plate chambers interspersed with iron.

The data sample used for the analyses contains 22.74~million 
$\BB$ pairs~\cite{ref:babar}, corresponding to 20.7\invfb\ taken
on the $\FourS$ resonance. In addition, 2.6\invfb\ of data taken
off-resonance have been used to validate the contribution to backgrounds 
resulting from $e^+ e^-$ annihilation into light \qqbar\ pairs. 
These data have all been processed with reconstruction software to determine 
the three-momenta and positions of charged tracks and the energies and positions 
of photons. Refined information on particle type from the various sub-detectors 
described above is also provided, and is used in particle identification 
algorithms in the analyses.

\section{Analysis Method}
\label{sec:Analysis}

\subsection{Dalitz Plot Decomposition for \pipipiz}
\label{sec:dpdecomp}
\begin{figure}[!htb]
\begin{center}
\includegraphics[height=9cm]{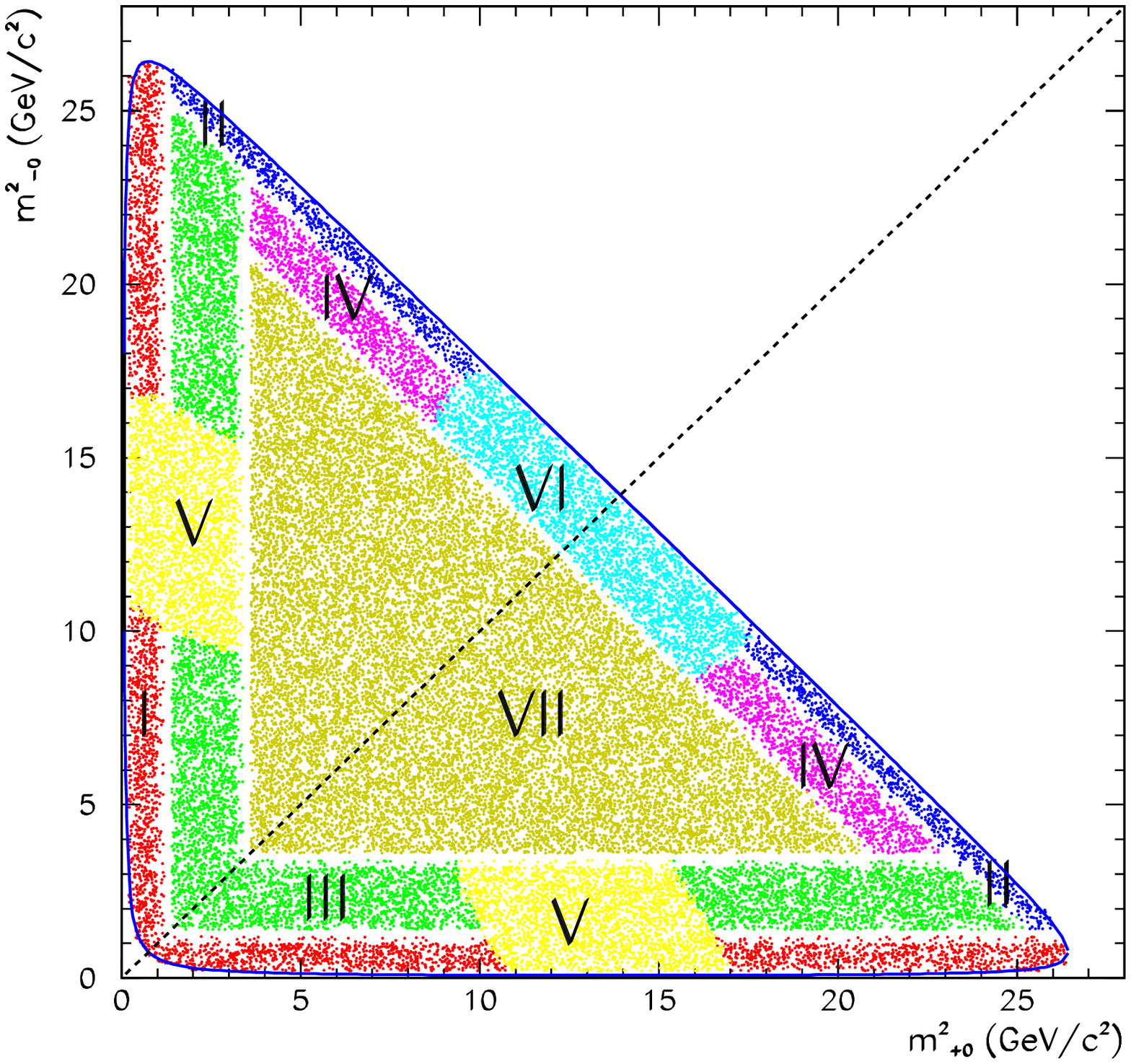}
\end{center}
\caption[Division of the Dalitz plot into regions.]
{The branching fraction into the Dalitz plot is measured in separate regions which 
are sensitive to the following modes: 
{\rm I}: \rhoppim,  {\rm II}: \rhozpiz,  {\rm III}: \rhopr, {\rm IV}: \rhoprz, 
{\rm V}: \sigmapm,  {\rm VI}: \sigmaz, {\rm VII}: \highmass. 
The dashed line along the diagonal represents the \CP-symmetry axis.}
\label{dpregions}
\end{figure}

Data for the \pppz\ final state can be represented on a Dalitz plot.
A complete picture of the Dalitz plot structure 
can be obtained only by performing an amplitude analysis, but the simplest
such analysis uses eight parameters~\cite{quinnsilva}, and would be 
difficult with the relatively low statistics currently available.
Previous analyses of such decays in this and other experiments 
\cite{cleo2000, belle2000, babar2000} 
have focused predominantly on resonant quasi two-body decays to $\rho(770)\pi$,
in which the data are localised in bands within the Dalitz plot.
We follow this approach, except that our current statistics are sufficient 
to allow also searches for higher resonances populating different bands
in the Dalitz plot.
We therefore sub-divide the Dalitz plot into several distinct regions, 
each of which is chosen to be sensitive primarily to a single resonance,
such as the $\rho(770)$, $\rho(1450)$ and $f_0(400-1200)$ 
(also called $\sigma$), and we perform an independent search in each. 
In addition, we consider the middle part of the Dalitz plot, which is 
sensitive to non-resonant decays.
We have found that the measured background level varies dramatically across 
the Dalitz plot and our approach allows the analysis for each 
resonance to be optimised independently, taking into account the local 
background level. The seven regions are indicated in Fig.~\ref{dpregions}.
The regions are defined using selection criteria 
for the invariant mass of $\pi\pi$-pair combinations, $m_{\pi\pi}$,
and the pair helicity angle, $\theta_H$, defined as the angle 
between the direction of one of the pions and the direction 
of the parent $B$ meson candidate, computed in the $\pi\pi$-pair rest frame. 
The region definitions are summarised in Table~\ref{regions}.
\begin{table}[!htb]
\caption{Regions of the \pppz\ Dalitz plot: the regions are kinematically-defined by 
the $m_{\pi\pi}$ and $|\cos{(\theta_H)}|$ selection criteria. Any possibly dominant 
resonance is noted for each region.} 
\begin{center}
\begin{tabular}{|c|c|c|}
\hline
Region     	& Putative dominant 	&  $m_{\pi\pi}$(\gevcc) and $|\cos{(\theta_H)}|$ selection \\
	   	& resonance		& \\
\hline
\hline
{\rm I}    	& $\rho^\pm(770)$	&  $0.52  < m_{\pi^+\pi^0}< 1.02 ~{\rm and}~ |\cos{(\theta^{+0}_H)}| > 0.25$ \\
           	&			&  $0.52  < m_{\pi^-\pi^0}< 1.02 ~{\rm and}~ |\cos{(\theta^{-0}_H)}| > 0.25$ \\
\hline
{\rm II}	& $\rho^0(770)$	&  $0.57  < m_{\pi^+\pi^-} < 1.02 ~{\rm and}~ |\cos{(\theta^\pm_H)}| > 0.3$ \\
           	&			&  $~{\rm and}~ m_{\pi^+\pi^0}> 1.2 ~{\rm and}~ m_{\pi^-\pi^0}> 1.2$ \\
\hline
{\rm III}	& $\rho^\pm(1450)$	& $m_{\pi^+\pi^0}~{\rm and}~m_{\pi^-\pi^0} ~{\rm and}~ m_{\pi^+\pi^-} > 1.2$ \\ 
           	&			& and (($m_{\pi^+\pi^0} < 1.8 ~{\rm and}~ |\cos{(\theta^{+0}_H)}| > 0.25$)  \\
           	&			& or   ($m_{\pi^-\pi^0} < 1.8 ~{\rm and}~ |\cos{(\theta^{-0}_H)}| > 0.25$)) \\
\hline
{\rm IV}	& $\rho^0(1450)$	& $1.2  < m_{\pi^+\pi^-} < 1.8 ~{\rm and}~ |\cos{\theta^{+-}_H}| > 0.3$ \\
           	&			& and $m_{\pi^+\pi^0}~{\rm and}~ m_{\pi^-\pi^0} > 1.9$ \\
\hline
{\rm V}		& charged scalar	& ($0.52  < m_{\pi^+\pi^0} < 1.8 ~{\rm and}~ |\cos{(\theta^{+0}_H)}| < 0.25$) \\
           	&			& or ($0.52  < m_{\pi^-\pi^0} < 1.8 ~{\rm and}~ |\cos{(\theta^{-0}_H)}| < 0.25$) \\
\hline
{\rm VI}	& $f^0(400-1200)$	&  $0.57  < m_{\pi^+\pi^-} < 1.8  ~{\rm and}~ |\cos{(\theta^{+-}_H)}| < 0.30$ \\
\hline
{\rm VII}	& high mass$(>1900)$	&  $m^2_{\pi^+\pi^0}~{\rm and}~m^2_{\pi^-\pi^0}~{\rm and}~m^2_{\pi^+\pi^-} > 0.13~m_B^2$ \\
\hline
\end{tabular}
\end{center}
\label{regions}
\end{table}
%

%

\subsection{Candidate Selection}
\label{sec:Selection}
Charged tracks are required to have a momentum less than 10\gevc\ and  
a transverse momentum greater than 0.1\gevc. They are required to have 
at least 20 hits in the DCH and to originate close to the beam-spot.
In addition, $\pipm$ candidates are selected as those charged tracks 
which are unlikely to be kaons, based on Cherenkov angle information 
from the DIRC combined with energy-loss information from the DCH.

Photon candidates are identified in the calorimeter as deposits of energy 
unassociated with charged tracks. \pizs\ are reconstructed by 
combining pairs of photon candidates and requiring that the invariant mass
of the resultant candidate is between 100\mevcc\ and 160\mevcc. 
Photon candidates used in \piz\ reconstruction are required to have a 
minimum energy of 30\mev
and a shower shape consistent with a photon. 

Reconstruction of $B$ candidates is accomplished by 
forming all combinations of
\pipipiz\ candidates in each event and applying a loose quality 
requirement of $\chi^2 < 200$ to the fitted
vertex for the $\pi^+\pi^-$ pair. The $B$ candidates 
are required to satisfy kinematic constraints appropriate for $B$ mesons. 
We use two kinematic variables~\cite{ref:babar} for this: 
$\mes = \sqrt{(\half s + \pvec_0\cdot \pvec_B)^2/E_0^2 - p_B^2}$,
where the subscripts $0$ and $B$ refer to the \epem\ system and the $B$ 
candidate, respectively; and $\Delta E = E_B^* - \sqrt{s}/2$, where $E_B^*$ 
is the \B\ candidate energy in the center-of-mass frame.
For signal events, the former has a value close to the $B$ meson mass 
and the latter should be close to zero.


The selection criteria in each region are optimized for sensitivity to the 
branching fraction. The criteria that are allowed to vary are the mass 
requirements on the $\pi^0$ and $\rho(770)$ candidates, $\cos{\theta_H}$, 
and the requirements on event shape discussed in the following section.

\subsection{Background Suppression and Characterisation}
\label{sec:background}
Charmless hadronic modes suffer very large amounts
of background from random combinations of tracks, mostly from light quark
and charm continuum production. Such backgrounds may be reduced by 
selection requirements on the event topologies computed in the 
$\FourS$ rest frame. We use $\cos{\theta_T}$, the cosine of the angle 
$\theta_T$ between the thrust axis of the $B$ meson decay and the 
thrust axis of the rest of the event. For continuum-related backgrounds, 
these two directions tend to be aligned because the reconstructed 
$B$ candidate daughters generally lie in the same jets as those 
in the rest of the event. By contrast, in $B$ events, the decay 
products from one $B$ meson are independent of those in the other, 
making the distribution of this angle isotropic. 
In consequence, requiring that this opening angle be significant provides a 
strong suppression of continuum backgrounds.

Other event shape variables also help to separate signal and background. 
We form a linear combination of 11 variables in a Fisher discriminant 
\xf\ \cite{Fisher}.
The coefficients for each variable are chosen to maximize the
separation between training samples of signal and background events. 
The variables contained in \xf\ are:

\begin{itemize}

\item the cosine of the angle between the $B$ momentum and the beam axis;
\item the cosine of the angle between the thrust axis of the $B$ candidate and the beam axis; and
\item the summed momentum,
in nine cones coaxial with the thrust axis of the $B$ candidate,
of all detected particles in the event that are not associated 
with the $B$ decay.
\end{itemize}

Despite the power of such topological variables to reduce the combinatorial 
backgrounds, most of the modes we have searched for continue to suffer 
significant background levels. Even after stringent selection criteria have 
been applied, it is necessary to make a background subtraction to isolate 
a signal or set an upper limit. In order to do this, the background in the 
signal region is estimated from the number of events in a sideband region, 
located near the signal region in the \mes-\DE\ plane, and extrapolated
into the signal region. The shape of the distribution of the background as 
a function of \mes\ is parameterised according to the ARGUS function 
\cite{argus1990}, and is measured using on-resonance data slightly 
displaced 
from the signal region in the \DE\ variable ($0.1 < |\DE\ | < 0.25$). 
We define \sigGSBratio\ to be the ratio of the number of candidates in the 
signal region to the number in the sideband region. 
Off-resonance data yields a consistent ratio and is 
averaged with the result measured on-resonance. We measure the value of 
\sigGSBratio\ independently for each region of the Dalitz plot, and we 
find that it is significantly dependent on it.

\subsection{Branching Fraction Analysis}
\label{sec:Physics}
The branching fractions are calculated according to 
\begin{equation}
{\cal B} = \frac{N_1-\sigGSBratio N_2}{N_{\BB} \times \epsilon}
\label{BReqn}
\end{equation}
where $N_1$ is the number of candidates in the signal region (SR) for
on-resonance data; $N_2$ is the number of candidates in on-resonance data 
observed in the sideband region (GSB), so that $\sigGSBratio N_2$ is the 
estimated number of background candidates in the signal region; 
$N_{\BB}$ is the number of $\BB$ pairs produced and $\epsilon$ is the signal 
efficiency. The GSB is specified by 
$5.21 < \mes < 5.25$\gevcc, $|\DE-\left<\DE\right>|<0.1$\gev, 
where $\left<\DE\right>$ is the mean value of \DE\ for $B$ meson decays 
found in calibration samples.

The numbers and distributions of candidates within the signal region 
remained unknown to us until all aspects of the analysis were finalized. The final 
selection criteria were chosen to maximize the sensitivity to the signal. 
Once chosen, neither the background description nor the cuts were changed.
This procedure was followed independently for each channel.

For the signal efficiency in Eq.~\ref{BReqn}, we used simulated signal
events and the same selection criteria as used for the data. The
efficiencies due to \piz\ reconstruction, particle identification, and \de\ 
selection criteria are determined from independent control samples
derived from the data.

We have adopted two slightly differing approaches to the definition of 
selection efficiencies in Eq.~\ref{BReqn}. For the \rhoppim\ and \rhozpiz\
measurements, we simulate the $\rho$ resonances according to a 
non-relativistic Breit-Wigner distribution within the phase-space and with
a $\cos^2{\theta_H}$ helicity distribution. For the non-resonant \pppz\
measurement, we simply assume a uniform distribution in phase-space,
and ignore interferences. In these cases, we take as the denominator in the
definition of the selection efficiency, the number of events generated in the 
full Dalitz plot, so that the loss of events outside our analysis cuts 
is compensated in Eq.~\ref{BReqn}. This provides a conventional
branching fraction, having assumed a signal distribution, and ignoring
interferences.
By contrast, for the measurements in the remainder of the Dalitz plot, 
we have no a priori model of the dynamical distribution of events in
phase space or of the feedthrough between different modes (other than
of \rhoppim, whose rate is known), so we simulate 
the events according to a uniform distribution in phase space. 
We note that this introduces no systematic error, provided that the 
efficiency is flat over the region. We have verified that this is
the case within the statistics available to us (the efficiency 
away from the kinematic boundaries, projected
onto either Dalitz plot axis is consistent with a linear dependence,
and the slope has been shown to be zero within errors of magnitude
$\pm 3\times 10^{-4}\gev^{-2}$). In these 
cases, we take as our denominator for the efficiency calculation, 
only those events generated within the boundary of the defined region.
In this way, events lying outside this boundary are not considered part of
the signal. 
For these latter ``topological'' branching fractions, the measurement
should be considered the total rate within the defined boundaries,
not just that associated with any single resonance. If a signal were
found, more work would be needed to determine the detailed dynamics.

\boldmath\subsection{Charge Asymmetry in \rhoppim}\unboldmath
\label{asymm}
In principle, there are four decay rates that we could measure
for the generic \rhoppim\ mode: we define
\begin{equation}
\Gamma(B^0\to \rho^+\pi^-)=\Gamma_{\rho\pi} ~;~
\Gamma(B^0\to \rho^-\pi^+)=\Gamma_{\pi\rho}
\label{twoRates}
\end{equation}
and their $CP$ conjugates
\begin{equation}
\Gamma(\Bzb\to \rho^-\pi^+)=\overline{\Gamma}_{\rho\pi} ~;~
\Gamma(\Bzb\to \rho^+\pi^-)=\overline{\Gamma}_{\pi\rho}
\label{CPconjgs}
\end{equation}
respectively. In the general, $CP$-violating case, all four are different, 
and there are two observable direct $CP$ violations:
\begin{eqnarray}
           \Delta_{\rho\pi}=(\Gamma_{\rho\pi}-\overline{\Gamma}_{\rho\pi})&\ne& 0 \nonumber \\
{\rm and~} \Delta_{\pi\rho}=(\Gamma_{\pi\rho}-\overline{\Gamma}_{\pi\rho})&\ne& 0.
\label{twoCPs}
\end{eqnarray}

In the present analysis, we have not implemented tagging, but we have 
divided our \rhoppim\ sample into two sub-samples,
containing respectively the events with $\rho^+\pi^-$ and those with
$\rho^-\pi^+$. Using these, we may form the charge asymmetry:
\begin{equation}
{\cal A}=\frac{N(\rho^+\pi^-)-N(\rho^-\pi^+)}{N(\rho^+\pi^-)+N(\rho^-\pi^+)}\simeq{\cal A_{\rm phys}}+{\cal A_{\rm det}}
\label{measuredAsymm}
\end{equation}
where ${\cal A_{\rm det}}$ is any detector-induced charge bias and
\begin{equation}
{\cal A_{\rm phys}}=
\frac{(\Gamma_{\rho\pi} + \overline{\Gamma}_{\pi\rho}) - (\overline{\Gamma}_{\rho\pi} + \Gamma_{\pi\rho})}
     {(\Gamma_{\rho\pi} + \overline{\Gamma}_{\pi\rho}) + (\overline{\Gamma}_{\rho\pi} + \Gamma_{\pi\rho})}.
\label{Aphys}
\end{equation}
The numerator of Eq.~\ref{Aphys} is simply the difference of the two 
direct $CP$ violations in Eq.~\ref{twoCPs}. If it is non-zero, it confirms 
direct $CP$-violation in at least one of the decays, 
Eqs.~\ref{twoRates} and \ref{CPconjgs}.
We may assume that systematics due to background subtraction and luminosity
cancel in the numerator of Eq.~\ref{measuredAsymm}.
Detector charge biases from tracking and from particle identification cuts were measured in 
an independent study with control samples of identified $D^+ \ra \KS \pip$ events. 
Single \pipm\ charge asymmetries were limited to $<2\%$, so we conservatively set 
a $\pm 2\%$ systematic uncertainty on our asymmetry measurement.

\section{Treatment of Experimental Uncertainties}
\label{sec:Systematics}

The systematic errors associated with branching fraction measurements arise from 
uncertainties in the background subtraction, in the overall signal efficiency 
and in the overall normalisation $N_{\BB}$.

In general, there are two contributions to the background subtraction:
continuum backgrounds and $B$-related backgrounds. The estimate of the 
continuum background is given as the product of the number of events in 
the GSB and the factor \sigGSBratio, which is the estimated ratio of the 
number of background events in the GSB to that in the signal region, 
and is ``measured'' with sidebands and off-resonance running as 
described above.
The number of events observed in the GSB has only a statistical error,
which is taken into account mode-by-mode.
The factor \sigGSBratio\ has a systematic error given by the error in the 
fitted ARGUS shape parameter, $\xi$. This is taken into account 
independently for each mode.

Possible sources of $B$-related backgrounds include events with 
low-multiplicity decays to charm, which have been excluded explicitly
by cuts in the Dalitz plot, and other charmless decays. 
The latter have been dealt with by counting the numbers found in the signal 
region in a dedicated charmless $B$ decay ``cocktail'' simulation, 
scaling them according to the relative luminosities in data and simulation
samples, and subtracting the appropriate numbers of events from the numbers 
in the signal region.
We include relevant statistical errors from the simulation sample,
and conservatively set the systematic error in this subtraction
to be equal to the number subtracted.

We measure the overall signal efficiency $\epsilon$ with signal 
Monte Carlo simulation, and some uncertainty
arises from limited signal mode statistics. The fractional error from this
source is 2.7\% in the \rhoppim\ mode. In the other modes, 
this error is negligible compared to other sources of uncertainty,
which are generally larger than the observed signals.
The accuracy of the simulation is subject to systematic uncertainties 
including the tracking efficiency, calorimetric shower efficiency, 
particle identification efficiency and the interaction of selection requirements with
the resolution accuracy of the simulation. At present the dominant
uncertainties are in the particle identification, $\pi^0$ reconstruction, and 
\de\ and event-shape modelling in the simulation.

A discrepancy between the measured track-finding efficiency, 
and that obtained in our simulation is taken into account with a 
momentum-dependent weight per track. The uncertainty in the determination
of these weights leads to an uncertainty in the tracking efficiency of 1.2\% 
per track, added coherently for both tracks used in the analyses. 
For $\pi^0$ reconstruction efficiency, we provide an energy-dependent 
weight, which is averaged over all accepted candidates in our signal 
simulation samples to provide a sample-dependent weight. We 
conservatively assign a systematic uncertainty of 5\% per $\pi^0$ 
to this weight.
The $\pi^{\pm}$ particle identification efficiency was measured in data with 
a $\pi^{\pm}$ control sample. It was found to be significantly smaller than
that predicted in the simulation, and this difference is taken into account 
with a factor $R^{corr} = 0.948 \pm 0.018$, per $\pi^{\pm}$, added coherently
for both pions.

A control sample of $B^+ \to \overline{D}^0\rho^+ \to K^+\pi^-\pi^+\pi^0$,
which has similar kinematics to our modes, has been used to show that the
simulation models the \de\ shape of such signals quite well. However,
due to statistical limitations in this method we ascribe a systematic
uncertainty of 5\% to the signal efficiency from our simulation, to account
for possible resolution differences on the order of 5\mev\ in \de.
Such uncertainties in the \mes\ selection efficiency are negligible by
comparison. For event shape cuts, the uncertainty on the efficiency has 
been taken as 5\%.

The overall normalisation, $N_{\BB}$, comes from a dedicated 
``$B$-counting'' study, which was made to determine
the number of $B$ mesons in the data samples. The final systematic error on 
this was determined to be 1.6\%.

The overall systematic uncertainty is the sum in quadrature of the 
contributions from all sources. Table \ref{syst_table} shows a summary
of sources and magnitudes of systematic errors.
\begin{table}[!ht]
\caption[Summary of Systematic Error Sources and Magnitudes]
{Summary of Systematic Error Sources and Magnitudes.}
\label{syst_table}
\begin{center}
{\small
\begin{tabular}{|l||c|c|}
\hline
Source of Uncertainty     & \multicolumn{2}{|c|}{Uncertainty} \\    
                          & \multicolumn{2}{|c|}{as \% of Result}  \\
			  \cline{2-3}
                          &   \rhoppim\       & Other modes \\
\hline
\hline
Background subtraction:   &  \multicolumn{2}{|c|}{}          \\
			  \cline{2-3}
${\cal A}$                & 5.2 &  $> 10$  \\
$B$-related Backgrounds     & 5.8 &  $> 20$  \\
\hline
sub-total                 & 7.8 &  $> 30$, dominant   \\
\hline
MC Statistics             & 2.7  &  negligible  \\
MC Efficiency Corrections & 10.5 &  negligible  \\
\hline
$B$-counting                & 1.6 &  1.6 \\
\hline
Total                     &14.5 &  $> 30$ \\
\hline
\hline
\end{tabular}

}
\end{center}
\end{table}
%

\section{Results}
\label{sec:Results}
Our preliminary measurements of branching fractions are summarized in Table 3.
\begin{table}[!htbp]
\label{tab:3pitable1}
\begin{center}
\begin{sideways}
\makebox[8.25in]{

\begin{tabular}{lcccc}
\multicolumn{5}{c}{Table 3: Results for the $3\pi$ final state analyses. We
aggregate the first two columns to obtain}\cr
\multicolumn{5}{c}{the branching fraction for $B^0 \rightarrow \rho^{\pm}\pi^{\mp}$ 
(the asymmetry is calculated independently.) } \cr\cr
\dbline

Quantity		 & $\Bz\to\rho^+\pi^-$	& $\Bz\to\rho^-\pi^+$	& $\rhozpiz$ 		& $\pppz(NR)$\cr
\sgline

$\cos \theta_T$	 	 & $<0.75$ 		& $<0.75$ 		& $<0.65$ 		& $<0.85 $\cr
Fisher                   & $<0.75$        	& $<0.75$        	& $<0.65$     		& $ < 1.1 $\cr
$\piz$ mass\gevcc	 & 0.110-0.160		& 0.110-0.160 		& 0.115-0.155 		& 0.115-0.155\cr

events in SR		 &  &  &  & \cr
~~On-res data		 & 121 			& 118			& 27 			& 41 \cr
~~Off-res data		 & 11 			& 12 			& 5 			& 3 \cr
events in GSB		 &  &  &  & \cr
~~On-res data		 & 590 			& 540 			& 142			& 362 \cr
~~Off-res data		 & 74			& 74 			& 24 			& 46 \cr
$\cal A$		 & $0.128\pm0.005$ 	& $0.128\pm0.005$ 	& $0.132\pm0.012$ 	& $0.116\pm0.006$ \cr
\qqbar\ BG in SR	 & $75.5\pm3.1\pm3.0$ 	& $69.1\pm3.0\pm2.7$ 	& $18.7\pm2.2\pm1.7$ 	& $42.0\pm3.1\pm2.0$ \cr
\bbbar\ BG in SR	 & $2.7\pm1.6\pm2.7$ 	& $2.7\pm1.6\pm2.7$ 	& $ 2.2\pm1.5\pm2.2$ 	& $3.2\pm1.8\pm3.2$ \cr
Signal			 & $42.8\pm11.5\pm4.0$ 	& $46.2\pm11.4\pm3.8$ 	& $6.1\pm5.8\pm2.8$	& $-4.2\pm7.3\pm3.8$ \cr
\sgline
Signal Efficiency	 & \multicolumn{2}{c}{$0.135\pm0.016$} 		& $0.074\pm0.009$ 	& $0.075\pm0.010$\cr
\sgline

Stat. signif. ($\sigma$)	& \multicolumn{2}{c}{5.01} 		& 0.96			& N/A \cr
$\cal B$ ($\times10^{-6}$)	& \multicolumn{2}{c}{$\BrRhoppim$}    	& $\BrRhozpiz$  	& N/A \cr
90\%\ CL limit ($\times10^{-6}$)& \multicolumn{2}{c}{N/A} 		& $<\UlRhozpiz$ 	& $<\UlPipipi$ \cr

\dbline
\end{tabular}

}
\end{sideways}
\end{center}
\end{table}

In calculating the upper limits, we have used the classical method 
outlined in \cite{pdg96}. In order to take account of the systematic
errors, we have reduced the background estimate and the efficiency by one 
standard deviation (systematic) before making the calculation.

The new value for the \rhoppim\ branching
fraction is more precise than previous measurements. 
Our upper limit for \rhozpiz\ is consistent with Standard Model 
expectations \cite{rho0pi0}. Our upper limit of $\UlPipipiVal$
for non-resonant \pipipiz\ decay ignores interference effects, an
approach similar to that adopted by CLEO \cite{CLEOnonres} for 
non-resonant $B^+$ decays to three charged pions.

In addition, we have made a preliminary measurment
of the $CP$ asymmetry defined in Eq.~(\ref{Aphys})
and find
\begin{equation}
{\cal A_{\rm phys}}=\AsRhoppimVal,
\end{equation}
consistent with zero. This does not however indicate that there 
is no direct $CP$ violation in \rhoppim\ decays, as the measured
quantity is the difference between two direct $CP$ violations, and we cannot 
exclude the possibility of a (coincidental) cancellation of the two.

\begin{figure}[!htb]
\begin{center}
\includegraphics[height=11cm]{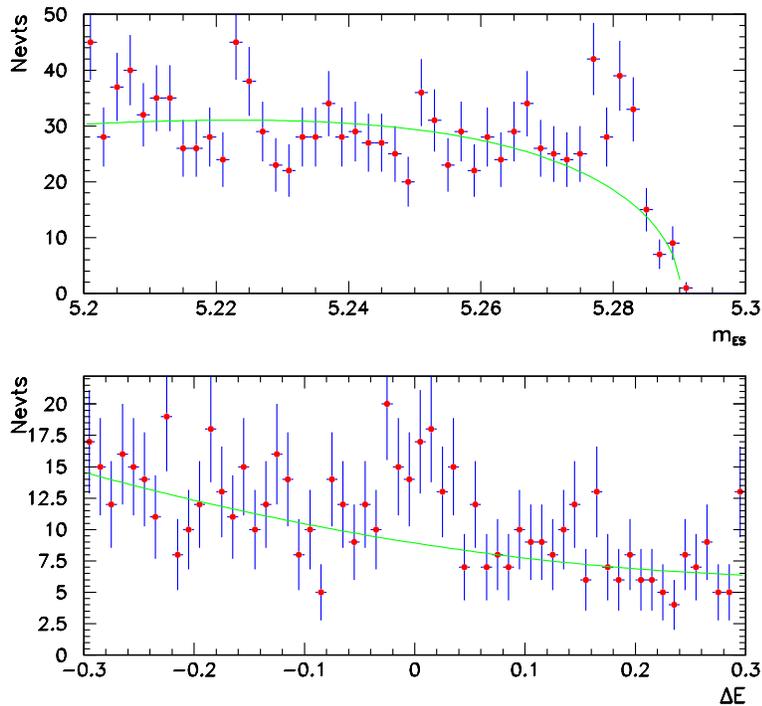}
\caption[]
{\mes\ and \de\ distributions for $B^0 \rightarrow \rho^{\pm}\pi^{\mp}$. The 
signal region requirement was made on the orthogonal variable in each case.}
\label{mesDePm}
\end{center}
\end{figure}

The signal for \rhoppim\ can be seen in the distributions of \mes\ and \DE\ 
shown in Fig.~\ref{mesDePm}.
The results for the topological branching fractions are given in Table 4. 
\begin{table}[!htbp]
\label{tab:3pitable2}
\begin{center}
\begin{sideways}
\makebox[8.25in]{

\begin{tabular}{lcccc}
\multicolumn{5}{c}{Table 4: Results for the $3\pi$ final state analyses.} \cr
\dbline

Quantity		& Region III  		& Region IV 		& Region V 		& Region VI \cr
\sgline

$\cos \theta_T$	 	& $<0.75$ 		& $<0.65$ 		& $<0.75$ 		& $<0.55$ \cr
Fisher                  & $<0.75$      		& $<0.70$      		& $<0.75$     		& $<0.40$ \cr
$\piz$ mass\gevcc	& 0.110-0.160 		& 0.115-0.155 		& 0.110-0.160  		& 0.115-0.155 \cr

events in SR		&  			&  			&  			&  	\cr
~~On-res data		& 75 			& 8 			& 44 			& 6  	\cr
~~Off-res data		& 4 			& 2 			& 0 			& 0 	\cr
events in GSB		&  			&  			&  			&   	\cr
~~On-res data		& 390 			& 80 			& 268 			& 22 	\cr
~~Off-res data		& 54 			& 6 			& 38 			& 2 	\cr

$\cal A$		& $0.131\pm0.006$ 	& $0.134\pm0.011$ 	& $0.132\pm0.009$ 	& $0.142\pm0.017$  \cr

\qqbar\ BG in SR	& $51.1\pm3.7\pm2.4$ 	& $10.7\pm1.7\pm0.8$ 	& $35.4\pm3.1\pm2.3$ 	& $3.1\pm0.9\pm0.4$ \cr

\bbbar\ BG in SR	& $6.5\pm2.5\pm6.5$ 	& $2.0\pm1.4\pm2.0$ 	& 0			& $3.2\pm1.8\pm3.2$ \cr

Signal			& $17.4\pm9.7\pm6.9$ 	& $-4.7\pm3.6\pm2.2$ 	& $8.6\pm7.3\pm2.3$ 	& $-0.3\pm3.2\pm3.2$ \cr
\sgline
Signal Efficiency	& $0.150\pm0.021$ 	& $0.087\pm0.018$ 	& $0.150\pm0.023$  	& $0.067\pm0.014$ \cr

\sgline

Stat. signif. ($\sigma$)& 1.83 			& N/A 			& 0.40			& N/A \cr
$\cal B$ ($\times10^{-6}$)& $\BrRhopr$ 		& N/A          		& $\BrSigmapm$ 		& N/A \cr
90\%\ CL limit ($\times10^{-6}$)& $<\UlRhopr$ 	& $<\UlRhoprz$ 		& $<\UlSigmapm$ 	& $<\UlSigmaz$ \cr

\dbline
\end{tabular}

}
\end{sideways}
\end{center}
\end{table}

The existence of a slight, though not statistically significant excess
in Region III could be a hint of structure due to higher resonances, 
such as the $\rho'^{\pm}(1450)$. This would have implications for the
measurement of $\alpha$ \cite{sophie}. Further data are necessary to
confirm whether this effect is real.

Region VI is sensitive to the scalar resonance $f_0(400-1200)$, proposed
as a significant contribution in this region \cite{sigma}. Our sensitivity 
has not yet reached that required to test this possibility.

\begin{figure}[!htbp]
\begin{center}
\includegraphics[height=16cm]{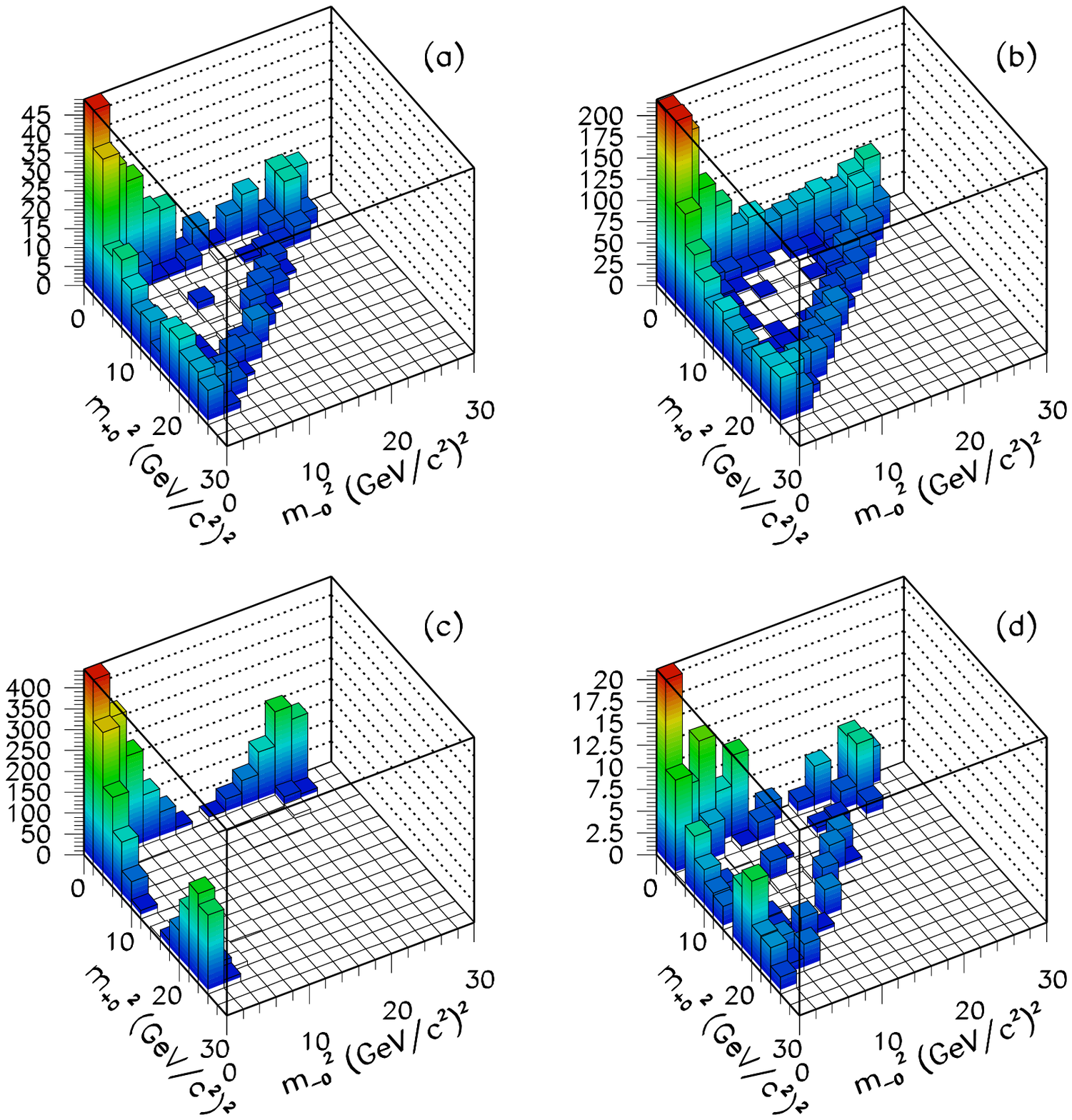}
\caption[Dalitz plots for \pppz\ after \rhoppim\ requirements.]
{Dalitz plots for \pppz\ after \rhoppim\ analysis requirements (except $m_{\pi\pi}$ 
and $|\cos{\theta_H}|)$: (a) data SR, (b) data GSB, (c) Monte Carlo SR, (d) 
background-subtracted data SR. }.
\label{dp3pidata}
\end{center}
\end{figure}

We can get an impression of the whole Dalitz plot from Fig.~\ref{dp3pidata},
which shows the full Dalitz plot area for various samples.
Fig.~\ref{dp3pidata} a) shows the distribution of the data in phase space
after the analysis cuts are applied, and Fig.~\ref{dp3pidata} b) shows 
the background distribution measured in the GSB. Figure \ref{dp3pidata} c)
shows an example Monte Carlo signal and Fig.~\ref{dp3pidata} d) shows
the background-subtracted signal. The helicity structure is clearly
visible along the two $\rho^{\pm}$-bands in the background-subtracted figure.
The peak near the origin (\ie\ low $\pi^0$ energy), even after background 
subtraction, indicates an efficiency enhancement in real signal events 
(also present in the simulated signal sample), in which events missing 
one low-energy photon 
due to acceptance effects are recovered by combinatorial substitution 
of a different photon. This effect significantly enhances the population
of signal events at low $\pi^0$ energies (only one candidate is
accepted per event).
%

\section{Conclusion}
\label{sec:Conclusion}
We have made a number of preliminary measurements of branching fractions 
into regions of the \pppz\ Dalitz plot. In particular, a new
measurement of the \rhoppim\ branching fraction, upper limits for the 
\rhozpiz\ mode and non-resonant \pppz, and four upper limit measurements 
of topological branching fractions into other regions of the Dalitz plot
have been reported.
We have also made the a preliminary measurement of the direct $CP$-violating 
asymmetry between the rates of untagged $\rho^+\pi^-$ and $\rho^-\pi^+$, 
finding no significant evidence for an effect.

\section{Acknowledgments}
\label{sec:Acknowledgments}


We are grateful for the 
extraordinary contributions of our \pep2\ colleagues in
achieving the excellent luminosity and machine conditions
that have made this work possible.
The collaborating institutions wish to thank 
SLAC for its support and the kind hospitality extended to them. 
This work is supported by the
US Department of Energy
and National Science Foundation, the
Natural Sciences and Engineering Research Council (Canada),
Institute of High Energy Physics (China), the
Commissariat \`a l'Energie Atomique and
Institut National de Physique Nucl\'eaire et de Physique des Particules
(France), the
Bundesministerium f\"ur Bildung und Forschung
(Germany), the
Istituto Nazionale di Fisica Nucleare (Italy),
the Research Council of Norway, the
Ministry of Science and Technology of the Russian Federation, and the
Particle Physics and Astronomy Research Council (United Kingdom). 
Individuals have received support from the Swiss 
National Science Foundation, the A. P. Sloan Foundation, 
the Research Corporation,
and the Alexander von Humboldt Foundation.


\end{document}